\documentclass[a4paper,12pt]{article}
\usepackage{amsmath}
\usepackage{amssymb}
\usepackage{fullpage}
\usepackage{verbatim}

\usepackage[T1,OT1]{fontenc}

\usepackage{graphicx}

\allowdisplaybreaks[3]

\usepackage[hang]{footmisc}
\usepackage[compress,square,numbers]{natbib}
\usepackage[colorlinks,linktocpage,linkcolor=blue,citecolor=blue,urlcolor=blue]{hyperref}

\renewcommand{\d}{\mathrm{d}}
\newcommand{\e}{\mathrm{e}}
\newcommand{\w}{\wedge}

\begin{document}

\numberwithin{equation}{section}

\thispagestyle{empty}

\vspace*{1cm}

\begin{center}

{\LARGE \bf Weakly Coupled de Sitter Vacua}

\vspace{0.5cm}

{\LARGE \bf with Fluxes and the Swampland}

\vspace{2cm}
{\large Daniel Junghans}\\

\vspace{1cm}
Institut f{\"{u}}r Theoretische Physik, Ruprecht-Karls-Universit{\"{a}}t Heidelberg,\\
Philosophenweg 19, 69120 Heidelberg, Germany\\

\vspace{1cm}
{\upshape\ttfamily junghans@thphys.uni-heidelberg.de}\\

\vspace{1.5cm}

\begin{abstract}
\noindent It was recently argued that the swampland distance conjecture rules out dS vacua at parametrically large field distances. We point out that this conclusion can in principle be avoided in the presence of large fluxes that are not bounded by a tadpole cancellation condition.
We then study this possibility in the concrete setting of classical type IIA flux compactifications with \mbox{(anti-)}O6-planes, \mbox{(anti-)}D6-branes and/or KK monopoles and show that, nonetheless, parametrically controlled dS vacua are strongly constrained. In particular, we find that such dS vacua are ruled out at parametrically large volume and/or parametrically small string coupling. We also find obstructions in the general case where the parametrically large field is an arbitrary field combination.
\end{abstract}

\end{center}

\newpage

\section{Introduction}

In recent years, a lot of effort has been devoted to constructing dS vacua in string theory. Starting with \cite{Kachru:2003aw}, many different scenarios and models have been proposed (see, e.g., \cite{Burgess:2003ic, Balasubramanian:2005zx, Rummel:2011cd, Louis:2012nb,Cicoli:2012fh, Cicoli:2013cha, Blaback:2013qza, Danielsson:2013rza, Rummel:2014raa, Braun:2015pza, Kallosh:2014oja, Marsh:2014nla, Guarino:2015gos, Retolaza:2015nvh, Dong:2010pm, Dodelson:2013iba, deCarlos:2009fq, deCarlos:2009qm, Dibitetto:2010rg, Danielsson:2012by, Blaback:2013ht, Damian:2013dq, Damian:2013dwa, Hassler:2014mla, Blaback:2015zra, Cicoli:2013rwa, Parameswaran:2006jh, Kounnas:2014gda, Achucarro:2015kja, Cicoli:2015ylx, Blumenhagen:2015kja, Blumenhagen:2015xpa, Gallego:2017dvd, Kobayashi:2017zfd, Antoniadis:2018hqy, Kallosh:2018nrk}). In order that all moduli are stabilized at a positive vacuum energy, these constructions require an intricate balance of several classical and quantum ingredients in the effective scalar potential. A possible concern with this approach is that the magnitude and moduli dependence of some of these ingredients and of possible corrections to them are not always known in full explicitness. While this may very well just be a technical issue, it has led some people to doubt the validity of the solutions. Furthermore, many no-go theorems rule out either the existence \cite{Gibbons:1984kp, deWit:1986xg, Maldacena:2000mw, Hertzberg:2007wc, Steinhardt:2008nk, Caviezel:2008tf, Flauger:2008ad, Danielsson:2009ff, Caviezel:2009tu, Wrase:2010ew, VanRiet:2011yc, Green:2011cn, Gautason:2012tb, Kutasov:2015eba, Quigley:2015jia, Andriot:2016xvq, Andriot:2017jhf, Andriot:2018ept} or the stability \cite{Covi:2008ea, Shiu:2011zt, Danielsson:2012et, Junghans:2016uvg, Junghans:2016abx} of dS extrema in various regions of the landscape. This has led to the interesting (although so far speculative) proposal that dS vacua might lie in the ``swampland'', i.e., that they are forbidden in quantum gravity (see, e.g., \cite{Brennan:2017rbf, Danielsson:2018ztv, Obied:2018sgi} and references therein and \cite{Cicoli:2018kdo, Akrami:2018ylq} for different perspectives).

This general idea was recently further specified in \cite{Obied:2018sgi} with the conjecture that an inequality $|\nabla V| \ge c V$ must hold everywhere in moduli space for some positive $\mathcal{O}(1)$ number $c$ (in Planck units).\footnote{For recent discussions of dark energy and inflation in this context, see, e.g., \cite{Colgain:2018wgk, Heisenberg:2018yae, Damian:2018tlf, Han:2018yrk, Ashoorioon:2018sqb, Heckman:2018mxl}.} If true, this inequality would in particular exclude all dS critical points, regardless of whether they are maxima or minima. Since there appear to be a number of counter-examples \cite{Denef:2018etk, Conlon:2018eyr,Roupec:2018mbn, Murayama:2018lie, Choi:2018rze, Hamaguchi:2018vtv, Olguin-Tejo:2018pfq, Blanco-Pillado:2018xyn}, several authors subsequently proposed a refined version of the conjecture \cite{Andriot:2018wzk, Garg:2018reu, Ooguri:2018wrx}, which states that the inequality should only hold unless the minimal eigenvalue of the Hessian is bounded from above, $\mathrm{min}\left(\nabla_i\nabla_j V\right) \le -c^\prime V$, where $c^\prime>0$ is another $\mathcal{O}(1)$ number (see also \cite{Dvali:2018fqu, Dvali:2018jhn, Garg:2018zdg} for other proposals).
Indeed, near no-scale points (which appear naturally in string theory), one can prove under mild assumptions that there is a universal tachyon at any dS critical point described by an F-term scalar potential \cite{Junghans:2016abx}. This is consistent with the above conjecture for $c^\prime=\frac{4}{3}$. Nevertheless, it is currently far from clear whether something like a no-dS conspiracy can be expected to hold in string theory \emph{in general}.

In the interesting recent paper \cite{Ooguri:2018wrx}, a step in this direction was taken by arguing that dS vacua are ruled out at parametrically large distances in moduli space. In order to show that, the authors used a general quantum gravity conjecture known as the swampland distance conjecture \cite{Ooguri:2006in}, together with an estimate of the dS entropy \cite{Gibbons:1977mu} based on the Bousso bound \cite{Bousso:1999xy}.\footnote{See also \cite{Moritz:2018sui} for an argument relating the dS conjecture to the weak gravity conjecture \cite{ArkaniHamed:2006dz} in a racetrack model and \cite{Dasgupta:2018rtp, Danielsson:2018qpa} for arguments against dS vacua at the quantum level.} The swampland distance conjecture, while not proven, has been extensively studied in the recent literature and verified in various examples \cite{Ooguri:2006in, Baume:2016psm, Klaewer:2016kiy, Valenzuela:2016yny, Blumenhagen:2017cxt, Palti:2017elp, Cicoli:2018tcq, Grimm:2018ohb, Heidenreich:2018kpg, Blumenhagen:2018nts, Grimm:2018cpv}. In its refined version \cite{Klaewer:2016kiy}, it states that a tower of exponentially light states with masses $m\sim \e^{-\alpha\Delta\phi}$ (for some $\alpha>0$) should appear as we move a super-Planckian distance $\Delta\phi \gtrsim 1$ in field space (see also \cite{Hebecker:2017lxm, Landete:2018kqf, Hebecker:2018fln} for some caveats). This implies in particular that such a tower must appear at parametrically large field distances, $\phi\to\infty$. The states in the tower were found in \cite{Ooguri:2018wrx} to lead to a dS entropy $S\sim \e^{a\phi}$ with $a>0$, which was then argued to imply a runaway in the scalar potential, $V\sim \e^{-a\phi}$. This result is reminiscent of an old argument due to Dine and Seiberg \cite{Dine:1985he}, where a runaway of this type was found for the case where $\phi$ is the dilaton. Another related argument, which bounds the potential from above by $\e^{-a\phi}$, was given in \cite{Hebecker:2018vxz} based on an estimate of the cutoff scale in theories satisfying the swampland distance conjecture. In the same paper, it was pointed out that a potential of the type $V\sim n(\phi) \e^{-a\phi}$ with some oscillating function $n(\phi)\sim\mathcal{O}(1)$ is a priori not excluded by the entropy or cutoff arguments and could lead to wiggles stabilizing the runaway. Such an oscillating potential may in particular appear for axionic fields.

In this note, we point out a different loophole to the argument of \cite{Ooguri:2018wrx} that exists in generic flux compactifications. In particular, we argue that, in the presence of fluxes that are unbounded by tadpole cancellation conditions, dS minima can in principle exist at parametrically large distances in moduli space without a runaway along the parametrically large field direction. We argue that this is possible without violating the swampland distance conjecture or the entropy bound, thus avoiding the no-go of \cite{Ooguri:2018wrx}.

One may wonder whether, in spite of this loophole, there is still a general obstruction to dS vacua at parametrically large field distances, perhaps due to some other reason unrelated to the above arguments. In fact, no parametrically controlled dS vacua have been constructed to date. While we do not present a general answer to this question, we analyze the possibility of parametrically controlled dS vacua in classical type IIA flux compactifications with \mbox{(anti-)}O6-planes, \mbox{(anti-)}D6-branes and/or KK monopoles.
The possibility of dS vacua in this general class of compactifications was studied before in many papers \cite{Hertzberg:2007wc, Silverstein:2007ac, Haque:2008jz, Caviezel:2008tf, Flauger:2008ad, Danielsson:2009ff, Danielsson:2010bc, Danielsson:2011au, Danielsson:2012et, Junghans:2016uvg, Junghans:2016abx, Roupec:2018mbn, Kallosh:2018nrk, Garg:2018zdg, Blaback:2018hdo}.
We find that dS vacua are ruled out in this setting at parametrically large volume and/or parametrically small string coupling (with all other fields fixed).
Our results are based on simple scaling arguments with respect to the moduli, similar to \cite{Hertzberg:2007wc}, and they do not assume the swampland distance conjecture.
We also study the case where the parametrically large field is an arbitrary direction in field space and find strong constraints in this case as well.

Let us stress that going to parametrically large distances in field space need not be required to achieve perturbative control. Indeed, a dS minimum at, say, a volume $\mathcal{V}=10$ and a string coupling $g_s=0.1$ can already be very well-controlled as long as one can argue that corrections that are subleading in the $g_s$ and $1/\mathcal{V}$ expansions have expansion coefficients $\ll \mathcal{O}(10)$. Nevertheless, it is interesting to contemplate a possible no-go against dS vacua at parametrically large field distances. For one thing, one may not always have full knowledge over the precise magnitude of subleading corrections but only know their scaling with respect to the expansion parameters. A parametric control over these corrections is then of course desirable. Second, in the spirit of the swampland program, it is worthwhile to understand possible universal properties of quantum gravity theories, including possible asymptotic properties. In this note, we will therefore only be concerned about statements at parametrically large field distances as in \cite{Ooguri:2018wrx}, keeping in mind that this does not rule out the existence of well-controlled dS vacua at moduli vevs which are large but not parametrically large. In the context of classical type IIA string theory, recent results \cite{Kallosh:2018nrk, Blaback:2018hdo} suggest that dS minima might indeed be possible in such a regime. We will briefly comment on this possibility at the end of the paper.

This paper is organized as follows. In section \ref{sec:ds}, we review the Dine-Seiberg runaway, its relation to the entropy argument of \cite{Ooguri:2018wrx} and how flux compactifications can avoid it.
Crucially, we show that this is consistent with the swampland distance conjecture and the exponential entropy scaling derived in \cite{Ooguri:2018wrx}. In section \ref{sec:IIA}, we study our loophole in the context of classical type IIA flux compactifications with \mbox{(anti-)}O6-planes, \mbox{(anti-)}D6-branes and/or KK monopoles. We find strong constraints on parametrically controlled dS vacua in this setting in spite of the presence of large fluxes. We conclude in section \ref{sec:concl} with suggestions for future research directions and some comments on dS vacua without parametric control.
\\

{\bf Note added:} After completion of this work, we were informed about \cite{Banlaki:2018ayh} which also discusses parametrically controlled dS vacua in type IIA string theory.

\section{Avoiding the Dine-Seiberg runaway}
\label{sec:ds}

Consider the scalar potential
\begin{equation}
V(\phi) = A \e^{-a\phi} + B \e^{-(a+b)\phi} + C \e^{-(a+c)\phi} + \ldots, \label{pot}
\end{equation}
where $\phi$ is a canonically normalized modulus and $a>0$, $c>b>0$, $A$, $B$, $C$ are coefficients. The dots stand for possible other terms with a higher suppression with respect to $\phi$. In the limit
\begin{equation}
\phi\to\infty, \qquad A,B,C= \text{const.}, \label{ds}
\end{equation}
only the first term in the potential survives,
\begin{equation}
V(\phi) \sim \e^{-a\phi}, \label{pot2}
\end{equation}
such that no minima $\partial_\phi V = 0$ are possible. In string compactifications, a potential of this type arises, for example, at weak string coupling (taking $\phi=-\ln(g_s)$) or at large volumes (taking $\phi\sim\ln\mathcal{V}$ with $\mathcal{V}=\int_X \d^6y\sqrt{g_6}$).
This is the well-known Dine-Seiberg problem \cite{Dine:1985he}. It was originally discussed for the case where $\phi$ is the dilaton but analogous problems exist for other moduli
(see, e.g., \cite{Denef:2008wq}). In particular, consider the volume modulus $\rho=\mathcal{V}^{1/3}$. In the limit $\rho\to\infty$, the potential is dominated by a leading term $V(\rho)\sim \rho^{-x}$ with $x>0$ such that no extrema can exist. Instead, the field either rolls towards a regime of small $\rho$ or there is a runaway to infinity. Dimensionally reducing the 10d Einstein-Hilbert term furthermore leads to a kinetic term of the form $\frac{(\partial\rho)^2}{\rho^2}$.
The canonically normalized modulus is therefore $\phi \sim \ln\rho$ such that we recover the exponential behavior \eqref{pot2}. In the following, we will refer to the limit \eqref{ds} as the Dine-Seiberg regime, regardless of whether $\phi$ is the dilaton or some other field.

In \cite{Ooguri:2018wrx}, the behavior \eqref{pot2} was derived for arbitrary fields (for the case $V>0$) without requiring a concrete string theory input. Instead, the authors used a combination of the swampland distance conjecture
and an entropy argument. They concluded that dS vacua are forbidden at asymptotic distances in moduli space. If true, this would in particular imply the intriguing result that dS vacua are forbidden at parametrically large volume and/or parametrically weak string coupling.
However, we argue that this conclusion can be avoided in the presence of flux.

Indeed, it is well-known that, in flux compactifications \cite{Douglas:2006es, Denef:2008wq}, the Dine-Seiberg runaway can be stabilized.
We will see below that this is consistent with the entropy scaling derived in \cite{Ooguri:2018wrx}.
However, let us first review how the Dine-Seiberg problem is avoided in the presence of flux. The key point is that the coefficients $A$, $B$, $C$ are then not constants but depend on the flux numbers, which can be adjusted such that terms with a different $\phi$-scaling balance each other out even at parametrically large $\phi$. For example, the potential \eqref{pot} admits dS minima at large $\phi$ for positive $A\sim \mathcal{O}(1)$, negative $B \sim \mathcal{O}(\e^{b\phi})$ and positive $C \sim \mathcal{O}(\e^{c\phi})$.\footnote{A simple example is $A=\frac{7}{2}$, $B=-4\lambda$, $C=\frac{3}{2}\lambda^2$, $a=b=1$, $c=2$, which yields a dS minimum at $\phi=\ln\lambda$ for arbitrarily large $\lambda$.} In order to find minima with parametrically large $\phi$, the large-$\phi$ limit has to be taken such that
\begin{equation}
\phi \to \infty, \qquad A, B \e^{-b\phi}, C \e^{-c\phi} = \text{const.} \label{flux}
\end{equation}
More generally than \eqref{pot}, we could also consider potentials with more than three terms that all scale differently with respect to $\phi$ but are of the same order due to the flux dependence of their coefficients. In this regime, it is a priori not a problem to find minima at arbitrarily large $\phi$.
It amounts to taking one or several fluxes parametrically large while approaching the asymptotic field distances. We will discuss this more explicitly in section \ref{sec:IIA} in the context of type IIA string theory. In the following, we will refer to this regime as the flux-balanced regime.

Typically, NSNS and RR fluxes are bounded by tadpole cancellation conditions of the form
\begin{equation}
\frac{1}{(2\pi\sqrt{\alpha^\prime})^{1+p}}\int_\Sigma H_3 \w F_p = Q^\text{loc},
\end{equation}
where the right-hand side stands for the charges of spacetime-filling branes and O-planes localized on a compact cycle $\Sigma$. This implies that the coefficients $B$, $C$ in \eqref{pot} cannot be made arbitrarily large such that, for sufficiently large $\phi$, we inevitably end up in the Dine-Seiberg regime where no minima exist.
However, there is no such bound whenever the fluxes arise from form fields whose legs are such that the $(3+p)$-form $H_3\w F_p$ vanishes on the compact space.\footnote{In the presence of geometric fluxes, which are related to the Ricci curvature of the internal space, there can be additional contributions to the tadpole conditions that need to be taken into account.} A well-known example are the type IIA AdS vacua of \cite{DeWolfe:2005uu}, where minima at parametrically large volume and weak string coupling were found by taking the $F_4$ flux large. While this particular compactification does not admit dS vacua \cite{Hertzberg:2007wc}, other compactifications have not been ruled out this way.\footnote{One can of course also find balanced potentials using quantum corrections (as, e.g., in \cite{Kachru:2003aw, Balasubramanian:2005zx}). However, it is not obvious whether such potentials can admit terms with unbounded coefficients $B$, $C$. The clearest example for an unbounded coefficient seems to be provided by a flux term without a tadpole condition, so this is what we will focus on in this paper.}

Let us now check that dS minima in the flux-balanced limit are consistent with the swampland distance conjecture. To see this, we briefly recall the argument of \cite{Ooguri:2018wrx}. The swampland distance conjecture states that, for large $\phi$, there should be an exponentially light tower of states with masses $m\sim \e^{-\alpha\phi}$ for some coefficient $\alpha>0$. These light states contribute to the dS entropy $S$. Using the Bousso bound and assuming that the light states saturate the dS entropy for asymptotic values of $\phi$,
it is then possible to show that
\begin{equation}
S \sim \e^{a\phi} \label{entr}
\end{equation}
for some other coefficient $a>0$.
Since the Gibbons-Hawking entropy of dS space \cite{Gibbons:1977mu} is inversely proportional to the vacuum energy, the authors of \cite{Ooguri:2018wrx} concluded that the field dependence of the scalar potential must be $V\sim \e^{-a\phi}$ off-shell, leading to the runaway behavior \eqref{pot2} described above.

However, this need not be the case in the flux-balanced regime \eqref{flux}. In particular, the entropy scaling \eqref{entr} constrains the magnitude, but not the off-shell field dependence, of the scalar potential to be $\sim \e^{-a\phi}$. Indeed, the entropy in the vicinity of the dS minimum described above is
\begin{equation}
S \sim \left( A \e^{-a\phi} + B \e^{-(a+b)\phi} + C \e^{-(a+c)\phi} + \ldots \right)^{-1} \sim \e^{a\phi}, \label{en}
\end{equation}
where the right-hand side follows from the fact that, as we argued before, a flux choice $ B \sim \e^{b\phi}$, $C\sim \e^{c\phi}$ is required at any minimum with parametrically large $\phi$. We thus recover the entropy scaling \eqref{entr} predicted by the swampland distance conjecture. We conclude that the existence of dS minima at parametrically large field values does not seem to be in conflict with the swampland distance conjecture, at least not based on a pure counting of the entropy.

In \cite{Ooguri:2018wrx}, the authors impose the additional assumption that the entropy must increase strictly monotonically in the limit $\phi\to\infty$. If this is true, the flux loophole is ruled out.
In particular, any displacement towards larger $\phi$ at fixed flux numbers would then lead to an increase of the entropy and, hence, a decrease of the potential, thus ruling out any minima.\footnote{Another loophole exploiting the possibility of a non-monotonic scalar potential is the ``periodic-wiggle'' loophole for axionic fields discussed in \cite{Hebecker:2018vxz}.}

\section{Weakly coupled dS vacua in type IIA}
\label{sec:IIA}

If dS vacua at parametrically large field distances are in the swampland as proposed in \cite{Ooguri:2018wrx}, then we face a puzzle: why should there be no dS vacua in the flux-balanced regime \eqref{flux}, even though it appears to avoid a runaway problem?
While we argued that the entropy scaling \eqref{entr} does not in general rule out such vacua, they may still be forbidden for some other reason.
In this section, we study this hypothesis in a class of string compactifications and indeed find obstructions to parametrically controlled dS vacua in spite of the loophole described above. 
Specifically, we consider classical type IIA flux compactifications with \mbox{(anti-)}O6-planes, \mbox{(anti-)}D6-branes and/or KK monopoles. This general class of compactifications has been studied before in many papers \cite{Hertzberg:2007wc, Silverstein:2007ac, Haque:2008jz, Caviezel:2008tf, Flauger:2008ad, Danielsson:2009ff, Danielsson:2010bc, Danielsson:2011au, Danielsson:2012et, Junghans:2016uvg, Junghans:2016abx, Roupec:2018mbn, Kallosh:2018nrk, Garg:2018zdg, Blaback:2018hdo}.
Recently, dS minima were found in this setup \cite{Kallosh:2018nrk, Blaback:2018hdo}, although so far at $\mathcal{O}(1)$ values for the volume and the dilaton and without properly quantized fluxes.\footnote{The dS solutions were furthermore obtained in the smeared approximation (see, e.g., \cite{Douglas:2010rt, Blaback:2010sj, Blaback:2011nz, Blaback:2011pn, Saracco:2012wc, McOrist:2012yc, Gautason:2015tig} for a discussion of this issue).} As we will see in the following, it is impossible to construct similar dS minima at parametrically large volume and/or parametrically small string coupling (with all other fields fixed). We also study the general case where the parametrically large field is an arbitrary direction in field space and find strong constraints in this case as well.

\subsection{Large volume and/or small string coupling}
\label{lv}

It is instructive to first discuss the case where the parametrically large field is the volume modulus, the dilaton or a combination of the two. This is arguably the most relevant case, as these two moduli are required to be large for perturbative control. We will later discuss a more general argument in section \ref{arb}, which applies to any direction in field space.

The general 4d scalar potential is \cite{Hertzberg:2007wc}
\begin{equation}
V(\tau,\rho) = - \frac{A_\text{O6/D6}}{\rho^{9/2}\tau^3} + \sum_{p=0,2,4,6}\frac{A_p}{\rho^{3+p}\tau^4} + \frac{A_3}{\rho^6\tau^2}
 + \frac{A_R}{\rho^4\tau^2}. \label{IIA}
\end{equation}
Here, we only display the dependence of the different terms with respect to the universal moduli
\begin{equation}
\tau = \frac{1}{g_s}, \qquad \rho= \mathcal{V}^{1/3},
\end{equation}
where $g_s=\e^\phi$ is the 10d dilaton and $\mathcal{V}=\int_X\d^6 y \sqrt{g_6}$ is the string-frame volume.\footnote{Our moduli are related to those in \cite{Hertzberg:2007wc} by $\rho=\rho_\text{\cite{Hertzberg:2007wc}}$ and $\tau=\rho^{-3/2}\tau_\text{\cite{Hertzberg:2007wc}}$.} The coefficients $A_i=A_i(\varphi_j,n_k)$ depend on the other, non-universal moduli (such as cycle volumes or axions) and the flux numbers. They are obtained by dimensionally reducing the 10d low-energy effective action of type IIA string theory.\footnote{In the presence of localized sources such as O-planes or D-branes, compactifications are typically warped. While the resulting subtleties in the 4d effective field theory are relatively well understood in type IIB/F-theory (see, e.g., \cite{DeWolfe:2002nn, Giddings:2005ff, Frey:2006wv, Koerber:2007xk, Shiu:2008ry, Douglas:2008jx, Frey:2008xw, Martucci:2009sf, Underwood:2010pm, Frey:2013bha, Martucci:2014ska, Grimm:2014efa, Grimm:2015mua, Martucci:2016pzt} and references therein), not much about this issue is known in type IIA. In the following, we ignore warping and assume that the 4d scalar potential is well-approximated by \eqref{IIA}.}
In particular, $A_p$ with $p$ even come from RR field strengths $|F_p|^2$ in the 10d action, $A_3$ comes from the NSNS field strength $|H_3|^2$ and $A_R$ comes from the internal Ricci scalar $R^{(6)}$. Note that $A_R$ may include the effects of so-called geometric fluxes and/or KK monopoles, which have the same scaling with respect to the universal moduli (see, e.g., \cite{Silverstein:2007ac, Caviezel:2009tu}).
The $A_\text{O6/D6}$-term comes from the DBI actions of \mbox{(anti-)}O6-planes and/or \mbox{(anti-)}D6-branes that wrap 3-cycles in the internal space. We refer to \cite{Hertzberg:2007wc} for the details of the dimensional reduction and explicit expressions. For the following discussion, only two properties of these terms will be important:
\begin{itemize}
\item All coefficients are non-negative,
\begin{equation}
A_\text{O6/D6}, A_R >0, \qquad A_3, A_p \ge 0.
\end{equation}
For $A_3$ and $A_p$, this follows because the NSNS and RR field strengths are sums of squares. $A_\text{O6/D6}$ could a priori be positive or negative, depending on whether it is dominated by the positive tension of branes ($A_\text{O6/D6}<0$) or the negative tension of O-planes ($A_\text{O6/D6}>0$). However, positivity here follows from the Maldacena-Nu\~{n}ez no-go theorem \cite{Maldacena:2000mw}, which states that a dS vacuum requires net negative tension, i.e., $A_\text{O6/D6}>0$. Finally, it is known that a dS vacuum in the present setting requires negative internal curvature \cite{Hertzberg:2007wc}. This corresponds to $A_R>0$.
\item The coefficients $A_3$, $A_p$ and $A_R$ can be parametrically large due to their dependence on flux numbers. For example, in the simplest case of a single 4-cycle $\Sigma$ threaded by $F_4$-flux, we have
\begin{equation}
A_4 \sim n^2, \qquad n = \frac{1}{(2\pi\sqrt{\alpha^\prime})^3}\int_\Sigma F_4.
\end{equation}
Here, $n$ can be a parametrically large number if $F_4$ is not bounded by a tadpole condition, as explained in section \ref{sec:ds}.
Analogously, the coefficients $A_3$, $A_p$ and $A_R$ can in general be parametrically large due to their dependence on NSNS/RR fluxes and geometric fluxes/KK monopoles, respectively. Crucially, it is \emph{not} possible for $A_\text{O6/D6}$ to be parametrically large. One might think that one could make use of the fact that $A_\text{O6/D6}$ depends on the brane and O-plane numbers. However, these are bounded due to the Maldacena-Nu\~{n}ez theorem by which the O-plane contribution must dominate over the brane contribution. The O-plane number is given by the number of fixed points of the orientifold involution and is therefore finite and fixed in a given compactification.
\end{itemize}
For later convenience, we also state the equations of motion for $\rho$ and $\tau$. From \eqref{IIA}, we find
\begin{align}
0 &= \rho\partial_\rho V = \frac{9}{2} \frac{A_\text{O6/D6}}{\rho^{9/2}\tau^3} - \sum_{p} (3+p)\frac{A_p}{\rho^{3+p}\tau^4} -6 \frac{A_3}{\rho^6\tau^2} -4 \frac{A_R}{\rho^4\tau^2}, \label{eom1} \\
0 &= \tau\partial_\tau V = 3 \frac{A_\text{O6/D6}}{\rho^{9/2}\tau^3} -4 \sum_{p}\frac{A_p}{\rho^{3+p}\tau^4} -2 \frac{A_3}{\rho^6\tau^2} -2 \frac{A_R}{\rho^4\tau^2}. \label{eom2}
\end{align}

Our claim is now that the potential \eqref{IIA} does not admit any dS vacua at parametrically weak string coupling ($\tau\to\infty$) and/or parametrically large volume ($\rho\to\infty$). Let us first consider the simple case where, as in the AdS vacua of \cite{DeWolfe:2005uu}, the unbounded flux is $F_4$. We therefore study the behavior of \eqref{IIA} in the limit $\lambda\to\infty$ with $A_4 \sim \lambda^2$ and $\rho\sim\lambda^r$, $\tau\sim\lambda^t$ for arbitrary exponents $r\ge0$, $t\ge0$. Note that we exclude negative exponents in order to avoid losing control. This yields the scalings
\begin{align}
& \frac{A_3}{\rho^6\tau^2} \sim \lambda^{-6r-2t}, && \frac{A_6}{\rho^9\tau^4} \sim \lambda^{-9r-4t}, && \frac{A_4}{\rho^7\tau^4} \sim \lambda^{2-7r-4t}, && \frac{A_2}{\rho^5\tau^4} \sim \lambda^{-5r-4t}, \notag \\ & \frac{A_0}{\rho^3\tau^4} \sim \lambda^{-3r-4t}, && \frac{A_\text{O6/D6}}{\rho^{9/2}\tau^3}\sim \lambda^{-9/2r-3t}, && \frac{A_R}{\rho^4\tau^2}\sim \lambda^{-4r-2t}. \label{scal}
\end{align}

Recall that, for a dS extremum to exist, we require both $A_\text{O6/D6}> 0$ and $A_R >0$. This corresponds to net negative tension \cite{Maldacena:2000mw} and negative internal curvature \cite{Hertzberg:2007wc}.
If the O6-plane term scales with a smaller power of $\lambda$ than the curvature term,
it will be diluted away at large $\lambda$ and thus effectively vanish. The potential \eqref{IIA} then only has positive terms at leading order and yields a runaway. Hence, it is clear that dS vacua are forbidden unless we choose $r$ and $t$ such that the O6-plane term scales with the same or a higher power of $\lambda$ compared to the curvature term. Using \eqref{scal}, it then follows
\begin{equation}
r\le-2t. \label{rt}
\end{equation}
This is incompatible with the requirements $r\ge0$ and $t\ge0$ unless $r=t=0$ such that dS vacua are ruled out whenever $\tau$ and/or $\rho$ are parametrically large.

What if one of the other fluxes or several different fluxes are unbounded by a tadpole condition?
We should then take into account the possibility of large fluxes $A_p \sim \lambda^{2c_p}$, $A_3 \sim \lambda^{2c_3}$ for arbitrary exponents $c_i\ge0$. However, it is clear that this does not affect the condition \eqref{rt}, which is due to the requirement that the $A_\text{O6/D6}$-term is at least of the order of the $A_R$-term in \eqref{scal}. To be as general as possible, we can also consider a parametrically large coefficient $A_R \sim \lambda^{2c_R}$ with $c_R\ge0$. This could happen, e.g., due to large geometric fluxes or a large number of KK monopoles.
We thus find
\begin{equation}
\frac{A_R}{\rho^4\tau^2}\sim \lambda^{2c_R-4r-2t}. \label{scalings}
\end{equation}
In order that the O6-plane term is not diluted away at large $\lambda$, we now require $r\le -2t-4c_R$. This is again incompatible with $r\ge0$ and $t\ge 0$ unless $r=t=c_R=0$.

The no-dS conclusion actually applies even more generally.
For example, we can consider compactifications including (anti-)NS5-branes. This yields an additional term
\begin{equation}
\frac{A_\text{NS5}}{\rho^5\tau^2}  \sim \lambda^{c_\text{NS5}-5r-2t}
\end{equation}
in the scalar potential, where we admit a large number of NS5-branes for generality, i.e., $A_\text{NS5}\sim\lambda^{c_\text{NS5}}$ with $c_\text{NS5}\ge0$. One can show that, in the presence of the NS5-branes, the requirement of negative internal curvature is dropped \cite{Hertzberg:2007wc} such that a priori we need not run into the above difficulties.
However, in order to allow dS minima for $\lambda\to\infty$, we now require that the O6-plane term is at least of the same order as the NS5-brane term. This is again due to our above argument that otherwise the O6-plane term would effectively vanish in the limit $\lambda\to\infty$. Hence, we have to consider $r\ge 2c_\text{NS5}+2t$. Substituting this into \eqref{scal}, we observe that 
the $A_0$ term now dominates over the $A_\text{O6/D6}$ term at large $\lambda$-values. This is again an obstruction to dS vacua unless we either set $r=t=c_\text{NS5}=0$ or $A_0=0$. The potential in the latter case is
\begin{equation}
V(\tau,\rho) = - \frac{A_\text{O6/D6}}{\rho^{9/2}\tau^3} + \frac{A_2}{\rho^5\tau^4} + \frac{A_4}{\rho^7\tau^4} + \frac{A_6}{\rho^9\tau^4} + \frac{A_3}{\rho^6\tau^2} + \frac{A_\text{NS5}}{\rho^5\tau^2}. \label{nof0}
\end{equation}
This satisfies $-\rho\partial_\rho V \ge \frac{9}{2}V$ and, hence, dS vacua are ruled out at parametrically large volume and/or parametrically weak string coupling even in the presence of NS5-branes.

\subsection{Arbitrary field combination}
\label{arb}

Let us now discuss the general situation where the parametrically large field is \emph{any} direction in field space. Indeed, one may wonder whether the above conclusion can be avoided if we exploit the fact that the coefficients $A_i$ can depend on non-universal moduli such as, e.g., 2-cycle or 3-cycle volumes. Instead of taking a limit where just a combination of $\rho$ and $\tau$ is parametrically large, one could try to also make some of these other moduli parametrically large or small in such a way that the scalings in \eqref{scal} change. As we will explain in the following section \ref{arb1}, a rather suggestive argument rules out this idea under broad conditions. Our argument is, however, not waterproof, and we will discuss possible loopholes in section \ref{arb2}.

\subsubsection{Argument}
\label{arb1}

Let us consider some field $\alpha$, which is an arbitrary combination of the universal and non-universal moduli. We assume that $\alpha$ parametrizes some path in field space such that we are at a large distance from a strongly coupled point (i.e., a point with $\mathcal{O}(1)$ values for the volume and the string coupling) in the limit $\alpha\to\infty$.

The scalar potential is of the general form
\begin{equation}
V(\alpha) = - \frac{\tilde A_{\text{O6/D6}}}{\alpha^x} + \sum_p \frac{\tilde A_p(\alpha)}{\alpha^{x}} + \frac{\tilde A_3(\alpha)}{\alpha^{x}} + \frac{\tilde A_R(\alpha)}{\alpha^{x}} \label{genpot}
\end{equation}
with $\tilde A_{\text{O6/D6}},\tilde A_R>0$ and $\tilde A_p,\tilde A_3\ge0$ as before.\footnote{We indicate with a tilde that we pulled out a different field combination than in section \ref{lv}.}
Here, without loss of generality, we defined the field $\alpha$ such that the O6-plane term scales like $\alpha^{-x}$ in the limit $\alpha\to\infty$. In the following, we want to focus on a regime of parametric control, which implies that the energy densities in \eqref{genpot} should become small in the limit $\alpha\to\infty$.\footnote{In particular, string-loop corrections are under parametric control if $g_s$ is parametrically small and $\alpha^\prime$ corrections are under parametric control if the string-frame energy densities in the 10d action are parametrically small. One can check that imposing either of the two implies that, after dimensional reduction and going to 4d Einstein frame, the energy densities in the 4d action are parametrically small.} The case $x<0$ is therefore excluded since then the energy densities diverge in the limit $\alpha\to\infty$ and we lose control. Similarly, we can neglect the case $x=0$ where the energy densities would approach constants instead of becoming small. For the remaining case $x>0$, we can then always set $x=1$ by a field redefinition.
Also note that the $\alpha$-dependence of the terms other than the first one is so far completely general. We have only pulled out a common factor $\alpha^{-1}$ for later convenience. As previously, we do not explicitly write out the dependence of the terms on fields other than $\alpha$ and on the flux numbers.

Recall now that the O6-plane term must not be subleading for large $\alpha$. Crucially, this term does not depend on any flux numbers and does therefore not contain any parametrically large numbers that are \emph{not} moduli. It is therefore not allowed to decay faster than the other terms in the limit $\alpha\to\infty$. This means that the functions $\tilde A_i$ are not allowed to diverge,
\begin{equation}
\lim_{\alpha\to\infty} \tilde A_i(\alpha) < \infty \qquad (i=0,2,3,4,6,R). \label{div}
\end{equation}
This would trivially be satisfied if all $\tilde A_i$ were constants in $\alpha$ but then the above potential would not yield any minima. However, the condition is also compatible with scalings such as, e.g., $\tilde A_i \sim \frac{n^2}{\alpha}$ or $\tilde A_i \sim n^2 \e^{-\alpha}$. As explained before, the corresponding terms can then still contribute to the leading-order potential in the limit of large $\alpha$ if the flux $n$ is large.

Consider first the simplest possible case where all terms have a power-law scaling with respect to $\alpha$, i.e., $\tilde A_i(\alpha)=\tilde B_i/\alpha^{y_i}$. This yields the potential
\begin{equation}
V(\alpha) = - \frac{\tilde A_{\text{O6/D6}}}{\alpha} + \sum_i \frac{\tilde B_i}{\alpha^{1+y_i}}, \qquad \tilde B_i, y_i\ge 0. \label{simple}
\end{equation}
Solving the $\alpha$ equation for $\tilde A_{\text{O6/D6}}$ and substituting this back into $V$, we find the on-shell potential
\begin{equation}
V(\alpha) = - \sum_i \frac{y_i\tilde B_i}{\alpha^{1+y_i}}. \label{ads}
\end{equation}
With $\tilde B_i\ge 0$ and $y_i\ge 0$, it follows that only AdS vacua are allowed. This is the essence of the problem with dS vacua at parametrically large field distances: the O6-plane term is the only negative term in the potential, and its coefficient cannot be made large. It must therefore be leading in the $\alpha$-expansion in order to contribute at large $\alpha$, which implies $\tilde B_i\ge 0$ and $y_i\ge 0$ for all $i$ and an absence of dS vacua.

Let us now analyze this problem for general $\tilde A_i$. The $\alpha$ equation is
\begin{equation}
0 = \frac{\tilde A_\text{O6/D6}}{\alpha} - \sum_{p}\frac{\tilde A_p}{\alpha} - \frac{\tilde A_3}{\alpha} - \frac{\tilde A_R}{\alpha} + \sum_{p}\tilde A_p^\prime + \tilde A_3^\prime + \tilde A_R^\prime, \label{alpha-eq}
\end{equation}
where $^\prime= \partial_\alpha$. Using \eqref{alpha-eq} in \eqref{genpot}, we can write the on-shell potential as
\begin{equation}
0 < V = \sum_p \tilde A_p^\prime + \tilde A_3^\prime + \tilde A_R^\prime, \label{xpot}
\end{equation}
where the right-hand side is strictly positive because $V>0$ at a dS vacuum.
Using furthermore \eqref{alpha-eq} in $V^{\prime\prime}$, we obtain the square of the $\alpha$-mass at the extremum, which needs to be positive:
\begin{equation}
0 < m_\alpha^2=V^{\prime\prime} = \sum_p \frac{\tilde A_p^{\prime\prime}}{\alpha} + \frac{\tilde A_3^{\prime\prime}}{\alpha} + \frac{\tilde A_R^{\prime\prime}}{\alpha}. \label{mass}
\end{equation}
Recall furthermore that all functions must satisfy $\tilde A_i\ge 0$. The key point is that these inequalities are difficult to satisfy while at the same time respecting that the functions should not diverge for large $\alpha$ (cf.~\eqref{div}). For example, as discussed above, it is impossible to satisfy \eqref{xpot} for $\tilde A_i \sim \alpha^{-y_i}$, $y_i> 0$ because then $\tilde A_i^\prime<0$. The same conclusion applies if $\tilde A_i$ is a sum of (positive) terms with such a scaling, or if we consider other decaying functions such as $\tilde A_i \sim \e^{-y_i\alpha}$ or $\tilde A_i \sim \ln(\alpha)^{-y_i}$. In any model with $\tilde A_i$'s of this type, parametrically controlled dS solutions are ruled out.\footnote{In some models, the equations of motion admit dS solutions that have parametrically large 3-cycle volumes and thus formally avoid this conclusion, see, e.g., \cite{Danielsson:2012et, Junghans:2016uvg}. However, these solutions either have parametrically small flux numbers or a parametrically large O6-plane number. This is not compatible with flux quantization and the fact that the number of orientifold fixed points is finite in a given compactification.}

Some more care is required when some of the $\tilde A_i$'s approach a constant at large $\alpha$, e.g., $\tilde A_i \sim n^2 f(\alpha)$ with $f(\alpha) = c_{1} - \frac{c_{2}}{\alpha} + \ldots$ and $c_1$, $c_2$ some $\mathcal{O}(1)$ coefficients. For $c_{1},c_{2}>0$, one can then satisfy $\tilde A_i > 0$ and $\tilde A_i^\prime > 0$ and thus potentially avoid the above arguments. Indeed, one checks that \eqref{xpot} and \eqref{mass} can both be satisfied if at least one of the $\tilde A_i$'s has a constant term, $c_1\neq 0$. However, these are only necessary conditions, and in fact one can show that such constant terms do not help to construct dS vacua either.\footnote{To see this, notice that a subleading term $\sim c_2$ can only have an effect in \eqref{genpot} at large $\alpha$ if the leading terms in the potential cancel, i.e., $- \frac{\tilde A_{\text{O6/D6}}}{\alpha} + \frac{n^2 c_{1}}{\alpha}=0$. Since there are no parametrically large numbers in the coefficient $\tilde A_{\text{O6/D6}}$, it then follows that the flux $n$ cannot be parametrically large either. The negative term $- \frac{n^2c_{2}}{\alpha^2}$ in \eqref{genpot} thus has an $\mathcal{O}(1)$ coefficient, which means that it must be the leading term in the large-$\alpha$ expansion in \eqref{genpot} or else is negligible at large $\alpha$ (subleading terms could only be non-negligible if multiplied by parametrically large fluxes). We thus arrive at a potential where only the leading term in the $\alpha$ expansion is negative and all other relevant terms are positive (because all $\tilde A_i$'s satisfy $\tilde A_i\ge 0$). One checks that such a potential cannot yield dS vacua, analogously to the situation discussed further above (e.g., around \eqref{simple} and below).}

We have thus seen that parametrically controlled dS vacua are ruled out under rather broad conditions, even if we allow the parametrically large field $\alpha$ to be an arbitrary field direction.

\subsubsection{Caveats}
\label{arb2}

In the remainder of this section, we will discuss several possible loopholes to these arguments.
In particular, we focussed so far on $\tilde A_i$'s which are dominated by one term with a unique scaling behavior at large $\alpha$.
However, it could be that the above problems are avoided if more general functions $\tilde A_i$ are allowed where several terms with naively different scalings compete at large $\alpha$ due to parametrically large fluxes $n_i$.  For example, consider
\begin{equation}
\tilde A_i \sim c_1\frac{n_1^2}{\alpha^{y}} - c_2\frac{n_1n_2}{\alpha^{y+1}} + c_3\frac{n_2^2}{\alpha^{y+2}} + \ldots. \label{ai}
\end{equation}
One checks that both \eqref{xpot} and \eqref{mass} can then be satisfied, and dS vacua can be obtained, if the fluxes $n_i$ and coefficients $c_i$ are such that the term in the middle is negative.

For the curvature term $\tilde A_R$, such a behavior is ruled out in many models
even though $\tilde A_R$ may be a complicated function of the moduli and the fluxes in the interior of the moduli space.
The reason is that, on many manifolds, the geometric fluxes cannot be made parametrically large because they are fixed by the geometry. At sufficiently large $\alpha$, $\tilde A_R$ is then inevitably dominated by a single term and thus satisfies a scaling law. For example, a well-studied class of negatively curved spaces are group manifolds, where the geometric fluxes are the structure constants of the group and, hence, fixed $\mathcal{O}(1)$ numbers (see, e.g., \cite{Caviezel:2008tf, Danielsson:2011au}). It would be interesting to study this further on other types of manifolds.

Similarly, $\tilde A_3$ and $\tilde A_p$ often have a simple scaling behavior at large $\alpha$ such that \eqref{ai} is ruled out.
In particular, they typically have the schematic structure (see, e.g., \cite{Caviezel:2008tf, Flauger:2008ad, Danielsson:2012et})
\begin{equation}
\tilde A_3, \tilde A_p \sim f(\varphi)\left( n_1 + n_2 b + n_3 b^2 + n_4 b^3 \right)^2 + \ldots, \label{axion}
\end{equation}
where $f(\varphi)$ is some a priori unknown function of all fields, $n_i$ are flux numbers and $b$ stands collectively for any axion. The dots denote further positive terms that have the same form as the displayed one. The number of terms appearing inside of the brackets depends on which of the $\tilde A_i$ we consider but this is irrelevant for our purpose. Consider now in particular the case where $\alpha$ is an arbitrary combination of non-axionic fields. The $\alpha$-dependence is then exclusively in the prefactor $f(\varphi)$. Hence, $\tilde A_3$ and $\tilde A_p$ are sums of positive terms which each have an overall scaling with respect to $\alpha$ in the limit $\alpha\to\infty$ and thus contribute negatively in \eqref{xpot}.

A well-studied class of compactifications of type IIA string theory is on SU(3)-structure manifolds. The form \eqref{axion} can then be shown to follow if the K\"{a}hler metric $K_{IJ}$ is diagonal. This is true for some simple group and coset spaces considered in previous scans of classical dS vacua \cite{Caviezel:2008tf, Flauger:2008ad, Danielsson:2009ff, Danielsson:2010bc, Danielsson:2011au, Danielsson:2012et, Junghans:2016uvg, Roupec:2018mbn, Kallosh:2018nrk, Garg:2018zdg, Blaback:2018hdo}. In general, however, the K\"{a}hler metric can have off-diagonal entries \cite{Caviezel:2008tf, Flauger:2008ad}, which would yield a structure $\tilde A_i \sim K^{IJ}(\varphi) \left(n_I+\ldots\right)\left(n_J+\ldots\right)$ instead of \eqref{axion}. The off-diagonal components of the metric can then produce $\alpha$-dependent terms that contribute with a minus sign as in \eqref{ai} and thus in principle evade our argument.\footnote{This conclusion cannot be avoided by rotating into a different basis with diagonal matrix $K^{IJ}$, as the $n_I$ vectors only correspond to integer, field-independent flux numbers in one distinguished basis.}
For example, in a simple two-field model and setting all axions to zero, we have $\tilde A_i \sim K^{11}(\varphi) n_1^2 + K^{22}(\varphi) n_2^2 + 2K^{12}(\varphi) n_1n_2$. For $K^{12}=0$, this is a sum of positive terms which depend on $\alpha$ only through the field-dependent prefactors $K^{11}$, $K^{22}$ such that the no-go argument of section \ref{arb1} applies. On the other hand, considering a non-zero off-diagonal term $K^{12}\sim -\frac{1}{\alpha^2}$ together with $K^{11}\sim \frac{1}{\alpha}$, $K^{22}\sim \frac{1}{\alpha^3}$, we find an expression as in \eqref{ai}, and it may be possible to avoid the no-go.
To see whether this works at large field distances, one would have to study negatively curved SU(3)-structure manifolds with sufficiently complicated moduli spaces, which we leave for future work.

Functions of the type \eqref{ai} may also occur if one allows that $\alpha$ involves axionic fields, e.g., taking $b\sim\alpha$ in \eqref{axion}. This apparent possibility to avoid the above problems is not surprising given the axionic shift symmetries. In the presence of fluxes, axions acquire a potential that breaks these symmetries, but there is still a combined shift symmetry acting on the flux numbers and the axions together. For example, in the simple case $n_3=n_4=0$, \eqref{axion} is invariant under shifts $b \to b+\delta b$, $n_1 \to n_1-n_2\delta b$.
Therefore, whenever \eqref{IIA} admits a dS minimum at any $\mathcal{O}(1)$ values of the moduli, we can always map it to another dS minimum with formally large axion vevs by applying a shift. However, the large field value here is just an artifact of the parametrization, and we could always undo it by applying the symmetry.
An analysis of whether dS vacua of \eqref{IIA} are excluded \emph{everywhere} on the moduli space is beyond the scope of the current work, and we conclude that axions do not help to evade our argument.

There are some other caveats worth mentioning. First, the issues pointed out in this section are ameliorated if the $\tilde A_i$ are allowed to diverge in the limit $\alpha\to\infty$.
A divergent $\tilde A_i$ may be possible if the function is proportional to a parameter which is not a modulus but can be taken parametrically \emph{small}. For example, $\tilde A_i \sim \epsilon \alpha^y$ with $y>0$ and $\epsilon\to 0$ could satisfy \eqref{xpot}, \eqref{mass} and still be of the same order as the O6-plane term in \eqref{genpot}. This is not possible for fluxes as they are quantized and can therefore at best be parametrically \emph{large}. However, one might imagine, for example, a parametrically small warp factor appearing in some of the terms. Such a possibility was already mentioned in \cite{Hertzberg:2007wc}.

Second, an implicit assumption in our arguments was that all factors in the terms of the scalar potential have a well-defined asymptotic behavior in the large-$\alpha$ limit. This would not be satisfied, for example, for periodic factors such as $\cos(\alpha)$ (see also \cite{Hebecker:2018vxz} for a discussion of such factors). While periodic functions are expected to occur along axionic field directions due to instanton corrections, it is not clear to us whether they could be relevant in the present setting where we considered a classical potential
and its dependence on the non-axionic fields.

Let us finally point out that, although our result rules out dS vacua under the discussed assumptions, we do not know whether this implies the conjectured \cite{Obied:2018sgi} property $|\nabla V|\ge cV$ for some $\mathcal{O}(1)$ number $c>0$. The reason is that our argument does not make any assumptions about the explicit form of the kinetic term and the potential for the field $\alpha$. Therefore, we do not know how the potential scales with the corresponding canonically normalized field. In compactifications of the type studied in \cite{Hertzberg:2007wc, Silverstein:2007ac, Haque:2008jz, Caviezel:2008tf, Flauger:2008ad, Danielsson:2009ff, Danielsson:2010bc, Danielsson:2011au, Danielsson:2012et, Junghans:2016uvg, Junghans:2016abx, Roupec:2018mbn, Kallosh:2018nrk, Garg:2018zdg, Blaback:2018hdo}, the kinetic terms of the (non-axionic) moduli are of the form $(\partial \alpha)^2/\alpha^2$ and the potential is power-law. Any slope in the potential is therefore exponential in the canonically normalized field. In that case, an absence of dS extrema indeed implies that $|\nabla V|\ge cV$. However, our above argument does not prove this in general.

\section{Conclusions}
\label{sec:concl}

In this note, we pointed out a possible loophole to the recent argument \cite{Ooguri:2018wrx} that dS minima are ruled out at parametrically large field distances. We argued that this conclusion can in principle be avoided in the presence of parametrically large flux. In particular, in the flux-balanced limit \eqref{flux}, dS vacua do not seem to be in contradiction with the swampland distance conjecture and the entropy scaling derived in \cite{Ooguri:2018wrx}.

We then analyzed whether our loophole allows to construct parametrically controlled dS vacua in the concrete setting of classical type IIA flux compactifications with \mbox{(anti-)}O6-planes, \mbox{(anti-)}D6-branes and/or KK monopoles. We explicitly showed that such dS vacua are ruled out at parametrically large volume and/or parametrically small string coupling, and we found strong constraints in the general case where the parametrically large field is an arbitrary field combination.

We leave it as an exercise for future work to check whether similar statements are true for compactifications with more general ingredients such as, e.g., \mbox{(anti-)}O4-planes, \mbox{(anti-)}D4-branes or quantum corrections. It would also be interesting to investigate compactifications in type IIB string theory as in \cite{Caviezel:2009tu}.
More ambitiously, one might attempt to formulate a general, model-independent argument (as in \cite{Ooguri:2018wrx}) against parametrically controlled dS vacua in the flux-balanced regime.

Finally, we stress again that our argument has nothing to say about dS vacua at large volume and weak string coupling per se, as long as they are not under parametric control.
For example, that dS minima might exist in the type IIA setting studied above is suggested by recent results that explain some of the difficulties in earlier attempts \cite{Danielsson:2012et, Junghans:2016uvg, Junghans:2016abx} and propose to avoid them using anti-D6-branes \cite{Kallosh:2018nrk} or KK monopoles \cite{Blaback:2018hdo}. Indeed, simple dS minima, although so far not under perturbative control (i.e., not at large $\rho$, $\tau$), were constructed in \cite{Kallosh:2018nrk, Blaback:2018hdo}.
It is at present not known whether there are similar minima also in the regime $\rho\gg1$, $\tau\gg1$. However, a simple scaling argument suggests that this might be the case.

To see this, recall the well-known fact that the type II string effective action at the classical two-derivative level has universal scaling symmetries \cite{Witten:1985xb, Burgess:1985zz}. This implies, for example, that the scalar potential \eqref{IIA} is invariant (up to an overall factor) under the rescaling
\begin{equation}
\rho \to \lambda^2\rho, \qquad \tau \to \lambda\tau
\end{equation}
if we rescale at the same time the coefficients such that
\begin{align}
& A_\text{O6/D6} \to \lambda^2 A_\text{O6/D6}, && A_p \to \lambda^{2p}A_p, && A_3 \to \lambda^4 A_3, && A_R \to A_R. \label{scal2}
\end{align}
Formally, any existing minimum in a given compactification can be mapped to a minimum at large volume and small coupling by the above rescaling. However, to reach large $\lambda$, we would need to rescale the flux numbers appearing in $A_i$ and the O6-plane number appearing in $A_\text{O6/D6}$ (note that a large number of D6-branes alone would not help because then we would get $A_\text{O6/D6} < 0$ instead of the required $A_\text{O6/D6} > 0$). Since the O6-plane number is fixed by the orientifold involution, we cannot take the limit $\lambda\to\infty$ in a given compactification and there is no parametric control.
However, as discussed above, parametric control need not be required in order to argue for a dS minimum. Indeed, the scalings \eqref{scal2} suggest that compactifications with sufficiently many orientifold fixed points (i.e., with $\lambda \gg1$) may admit minima analogous to those in \cite{Kallosh:2018nrk, Blaback:2018hdo} at large volume and weak string coupling.  It would be very interesting to study the mechanisms found in \cite{Kallosh:2018nrk, Blaback:2018hdo} further on such orientifolds and check whether there are instabilities \cite{Danielsson:2016cit} or other obstructions to dS vacua there.

\section*{Acknowledgments}

I would like to thank Arthur Hebecker, Gary Shiu, Pablo Soler, Thomas Van Riet and Timm Wrase for helpful discussions. I am supported in part by the DFG Transregional Collaborative Research Centre TRR 33 ``The Dark Universe''.

\bibliographystyle{utphys}
\bibliography{groups}

\end{document}